# Vibration-Induced Conductivity Fluctuation Measurement for Soil Bulk Density Analysis

Andrea Sz. Kishné*[a], Cristine L.S. Morgan[a], Hung-Chih Chang[b], Laszlo B. Kish[b]
[a]Dept. of Soil and Crop Sciences, Texas A&M Univ., College Station, TX, 77843-2474, USA;
[b]Dept. of Electrical and Computer Eng., Texas A&M Univ., College Station, TX, 77843-3128, USA

## ABSTRACT

Soil bulk density affects water storage, water and nutrient movement, and plant root activity in the soil profile. Its measurement is difficult in field conditions. Vibration-induced conductivity fluctuation was investigated to quantify soil bulk density with possible field applications in the future. The AC electrical conductivity of soil was measured using a pair of blade-like electrodes while exposing the soil to periodic vibration. The blades were positioned longitudinally and transversally to the direction of the induced vibration to enable the calculation of a normalized index. The normalized index was expected to provide data independent from the vibration strength and to reduce the effect of soil salinity and water content. The experiment was conducted on natural and salinized fine sand at two moisture conditions and four bulk densities. The blade-shaped electrodes improved electrode-soil contact compared to cylindrical electrodes, and thereby, reduced measurement noise. Simulations on a simplified resistor lattice indicate that the transversal effect increases as soil bulk density decreases. Measurement of dry sand showed a negative correlation between the normalized conductivity fluctuation and soil bulk density for both longitudinal and transversal settings. The decrease in the transversal signal was smaller than expected. The wet natural and salinized soils performed very similarly as hypothesized, but their normalized VICOF response was not significant to bulk density changes. This lack of sensitivity might be attributed to the heavy electrodes and/or the specific vibration method used. The effects of electrode material, vibration method and soil properties on the experiment need further study.

*akishne@ag.tamu.edu; phone 1 979 845 2541; fax 1 979 845 0456

## 1. INTRODUCTION

Soil bulk density, defined as oven dry mass per unit volume of soil, is an important soil property which has applications to many soil studies. For example, soil bulk density is used in estimating water budgets, nutrient availability, and soil carbon sequestration in the plant root zone. The problem is that soil bulk density varies across landscapes and its measurement is time consuming and difficult in field conditions. Methods for measuring bulk density include traditional volumetric ring, paraffin sealed clod, and gamma ray attenuation techniques [1, 2]. The traditional volumetric ring measurement is a destructive measurement. The nuclear method requires water content determination independently to enable the calculation of bulk density and access tube installed in the field. Time-domain reflectometry (TDR) technique is developed to measure apparent dielectric constant and bulk electrical conductivity simultaneously in-situ determination of soil water content and bulk density [3]. The disadvantage of TDR is the fading measurable signal in wet saline soils. A novel method for quantifying soil bulk density that is potentially applicable in field conditions is based on measuring vibration-induced conductivity fluctuation (VICOF) [4]. The AC electrical conductivity is measured with and without exposing the soil to gentle vibration--without compacting it. The vibration-induced elastic soil density fluctuation generates a corresponding conductivity fluctuation (dRs). The current response is measured at the sum and difference of the mean AC and the double of vibration frequency. The conductivity fluctuation is normalized to the soil resistance (Rs). The normalized VICOF (dRs/Rs) is inherently independent of soil resistance, which is strongly influenced by moisture and salinity [5]. Thus normalized VICOF is expected to depend only on soil porosity and its related mechano-electrical properties. It has been shown for clay and sand soils at two moisture conditions that the conductivity fluctuation of soil and its normalized value to the soil conductivity are strongly related.

The bulk density demonstrated inverse curvilinear relationship to the normalized VICOF in an exploratory experiment [6].

The objective of this study is two-fold: 1. To improve the normalized VICOF signal to noise ratio by using blade shape electrodes; 2. To reduce the effect of moisture and salinity on the determination of bulk density by positioning the blade electrodes longitudinally and transversally to the direction of the induced vibration, and calculating the ratio of the transversal normalized ($dR_{s,T}/R_s$) and longitudinal ($dR_{s,L}/R_s$) normalized conductivity fluctuations.

# 2. METHODOLODY AND DATA

## 2.1. Measurements

The measurement circuitry is shown in Figure 1. It was the same arrangement as the one used by [4].

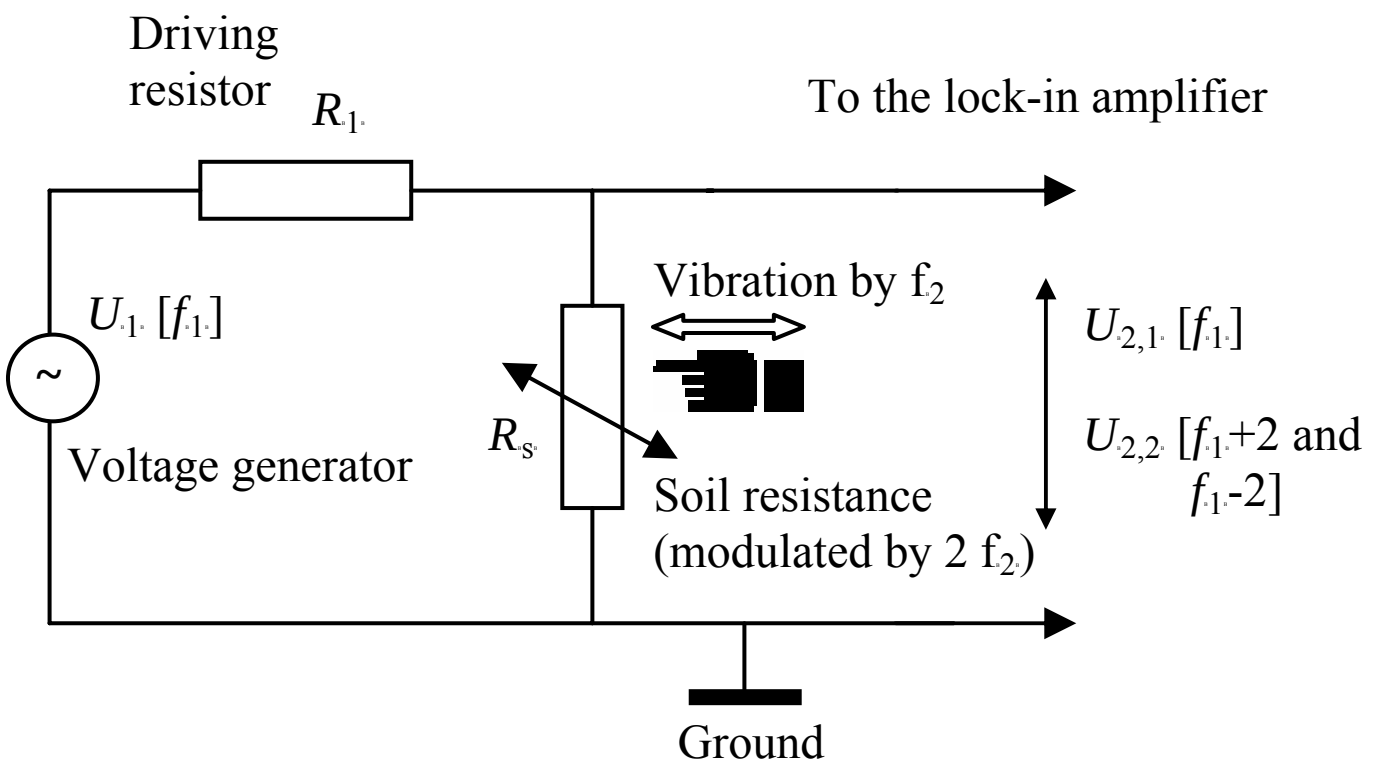

Fig. 1. Measurement circuitry. The AC current ($U_1$) was provided by a lock-in amplifier at $f_1$ = 1 kHz. Rs represents the soil sample. The periodic vibration frequency ($f_2$) was 60 Hz. The driving resistor ($R_1$) was selected so that the current through the sample was approximately 0.5 x $U_1$.

The electrode pair was placed in the soil 4.3 cm deep from the soil surface and 1.5 cm apart in the center (Fig. 2). A guiding template assured the distance and was removed before measurement. The electrode material was stainless steel grade 304. The length, width and thickness of electrodes were 7, 2, and 0.15 cm, respectively. The cutting edge of each blade was sharpened.

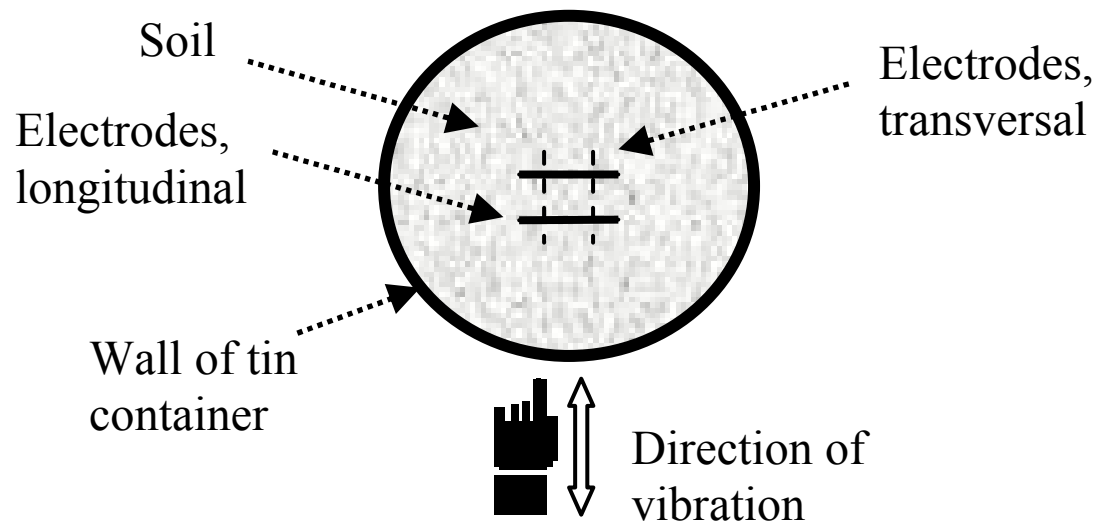

Fig. 2. Set-up arrangement of the measurement. The current flow of the electrodes is positioned longitudinally and transversally to the direction of vibration. The sample holder was placed on a floating top of an antivibration table providing that the vibration was horizontal in a well-defined direction.

At each longitudinal and transversal position, $U_1$, $U_{2,1}$ were measured at 1 kHz frequency, averaged, and recorded for 10 seconds at the beginning and end of a measurement set. The sample averages were used in further calculations. The background $U_{2.2,0}$ and vibration-induced $U_{2.2,v}$ signal were measured at 1.12 kHz minimum of 5 times for 30 seconds alternating the longitudinal and transversal positions and the 30 second averages were recorded. The recording started after the fluctuation of signals became stable.

The AC resistance of the soil sample ($R_s$) is calculated in the classical way according to (1).

$$R_s = R_1 \frac{U_{2,1}}{U_1 - U_{2,1}} \ . \qquad (1)$$

Supposing small modulation, we estimated the conductance modulation ($dR_s$) induced by the periodic vibration and the normalized dRs/Rs from the voltage modulation ($U_{2,2}$) according to the following equations (2) and (3). The detailed deliverance of these equations can be found at [4].

$$dR_s = 2R_1 \frac{U_{2,2}}{U_1 - U_{2,1}} (1 + \frac{U_{2,1}}{U_1 - U_{2,1}}) \ , \qquad (2)$$

and

$$\frac{dR_s}{R_s} = 2\frac{U_{2,2}}{U_{2,1}}\left(1 + \frac{U_{2,1}}{U_1 - U_{2,1}}\right) \ . \qquad (3)$$

The voltage modulation ($U_{2,2}$) is the difference of the signal ($U_{2,2,v}$) induced by the vibration and the background voltage ($U_{2,2,0}$) measured without vibration. The magnitude of $U_{2,2,0}$ and $U_{2,2,v}$ is in the order of microV. For the calculation of $U_{2,2}$, the simple difference of $U_{2,2,v}$ and $U_{2,2,0}$ was used in [4, 6], but in the current study, we applied a more precise expression according to equation (4).

$$U_{2,2} = \sqrt{U_{2,2,v}{}^2 - U_{2,2,0}{}^2} \ . \qquad (4)$$

## 2.2. Computer simulation of the transversal effect

We run computer simulations with a simplified model of a two-dimensional 10x10 square lattice of uniform resistors, see Figure 4 (a). We induced defects by cutting out resistors and measured the defect density by the filling factor *p*. This p factor was 100% for the original perfect lattice; *p*=0% for the lattice with no remaining resistor and, for the infinite lattice, the percolation limit (when the macroscopic resistance diverges to infinity) is *p*=50%. After determining the macroscopic resistance of the system in the longitudinal and the transversal directions at various *p* values, and the microscopic current densities, we used the Cohn-Thellegen theorem to calculate the average sensitivity of the system for the removal of an extra resistor in the longitudinal or the transversal direction. This simulation did not take into account the mechanical side of the effect: how the pressure fluctuations modulate the microscopic resistance due to given periodic acceleration.

## 2.3. Soil samples and treatments

The soil samples used in the experiment were from the A-horizon of a fine sand, hydrophobe soil from Caldwell, Texas, with 0.7 % organic carbon. It was non-saline as indicated by its 1.5 dS $m^{-1}$ electrical conductivity measured in saturated paste. The air-dried, crushed soil was passed through a 2-mm sieve and mixed thoroughly. The sample was divided for wetting by distilled water and by electrolyte solution. The wet moisture condition was prepared for .045 g $g^{-1}$ water content on a dry-mass basis corresponding to -33 J $kg^{-1}$ matric potential. The dry soil was wetted to .026 g $g^{-1}$ water content on a dry-mass basis corresponding to -1500 J $kg^{-1}$ matric potential. The bulk soils were mixed thoroughly regularly and equilibrated in closed bags for minimum of two months. The amounts of soil for four replicates of four bulk density were treated separately. The actual moisture of each sample was sampled and determined after each VICOF measurement, and it averaged .041 g $kg^{-1}$ on a dry-mass basis (SD=.0006) for wet soils including the saline samples, and .0224 g $g^{-1}$ on a dry-mass basis (SD=.0004) for dry moisture conditions. For salinization, the solution was 45 meq $l^{-}$ [1] concentrate with NaCl and $CaCl_2$ (50-50 %), and was added to the air dry soil to wet it to -33 J $kg^{-1}$ matric potential condition. Its EC was 15.8 dS $m^{-1}$ measured in saturated paste.

Samples were layered as evenly as possible in painted tin sample holders (9.7 cm diameter and 6.3 cm height) just before the VICOF measurements. Between layering, the compacted surface was scratched randomly to improve the connection to the next soil layer. Four compaction levels were prepared by using a 1 kg weight or a 3 kg manual soil compactor with a wood adaptor fitting in the sample holder from heights of 2 cm, 5 cm and 13 cm, 2 or 5 times depending on the compaction level. The bulk density values resulted from this preparation are listed in Table 1.

Table 1. Bulk density calculated from the sample weight and the volume of container. Average value of sample replicates is listed in the table with the standard deviation in parenthesis.

| **Compaction level** | **Average bulk density** | | |
|---|---|---|---|
| | Dry sand | Wet sand | |
| | non-saline | non-saline | saline |
| | ----------------------------g $cm^{-3}$-------------------------- | | |
| 1 | 1.25 (0.00) | 1.13 (0.01) | 1.12 (0.00) |
| 2 | 1.31 (0.00) | 1.20 (0.00) | 1.19 (0.00) |
| 3 | 1.36 (0.01) | 1.30 (0.01) | 1.30 (0.01) |
| 4 | 1.54 (0.00) | 1.49 (0.01) | 1.48 (0.01) |

After the VICOF measurement, the signals were evaluated and filtered for relaxation of $dR_s$ and possible contact problem indicated by the large scatter of $dR_s$ over 2 times the SD of signals. Out of 38 pairs of longitudinal and transversal measurements, 2 samples were considered as outliers, 2 longitudinal and 4 transversal measurement sets were taken out because of relaxation that was unseen from the raw data.

# 3. RESULTS

## 3.1. Set-up noise

The applied electrodes and their geometric arrangement, voltages, frequencies, the electrode-soil interface, and the vibrator-soil connection all contribute to the set-up noise of the measurement. A simple comparison of the currently used blade-shape electrode pair and a cylindrical electrode used in the previous experiment [4, 6] is presented in Figure 3. The cylindrical electrode was placed in the soil in four locations, while the other electrode was always clamped to the sample holder diagonally to the vibrator. The sample holder was turned 90° but the electrodes were positioned the same way relative to the vibration direction. At each location, $U_{2,1}$, $U_{2,2,0}$, and $U_{2,2,v}$ were measured for 1.5 minutes and the average was recorded. Representative sample was just measured for the least and most dense conditions. At the current project, all samples were measured for characterizing this noise and the standard error of sample averages are presented as error bars in Figure 3.

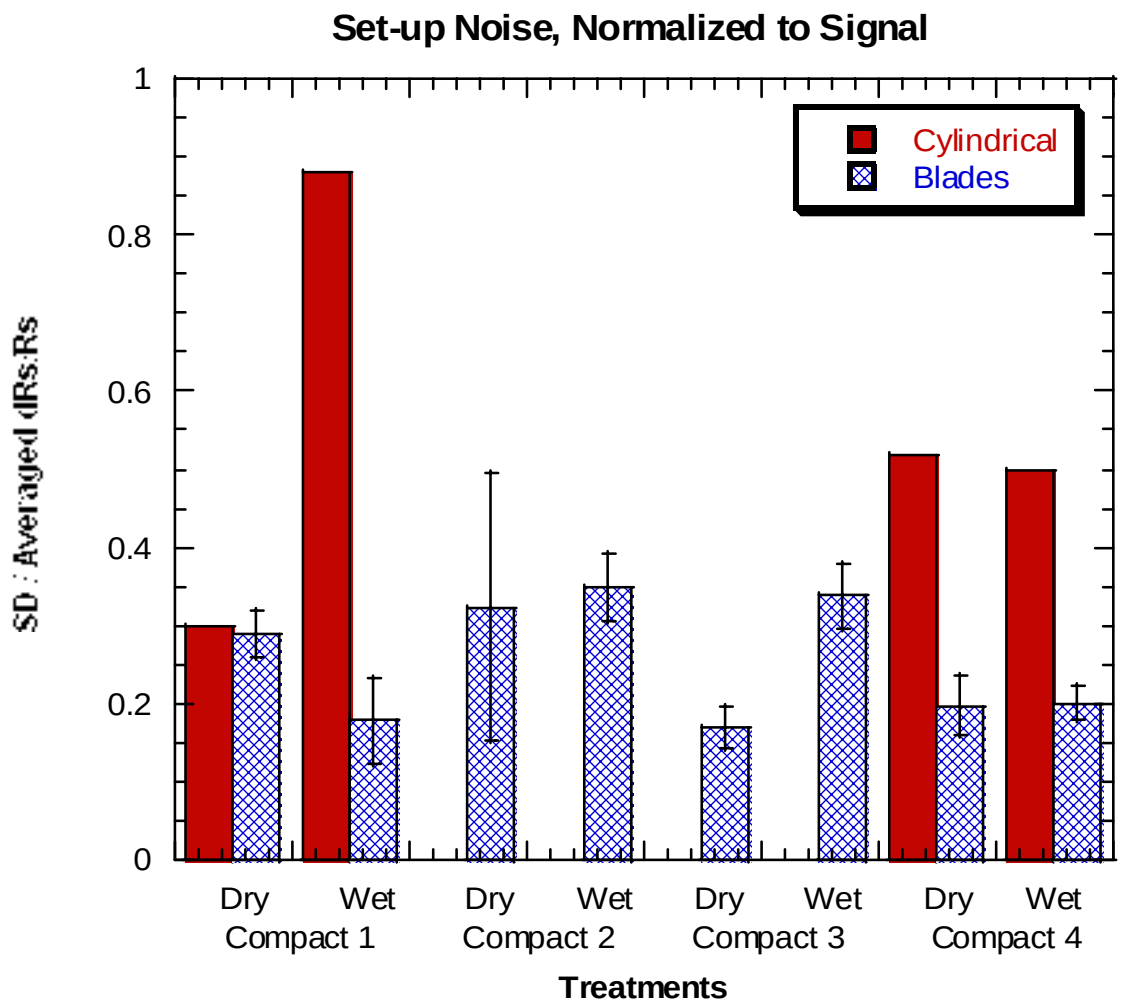

Fig.3. Variability of the VICOF measurement is characterized at a specific location by the ratio of the standard deviation relative to the signal, the averaged normalized VICOF of a sample. The error bar shows the standard error of sample variation where more than one sample was measured.

The comparison indicates that the blades improved the soil-electrode connection. Although there was only one sample per treatment measured for the cylindrical electrodes, the blades seem to reduce this noise at least to half for the wet and least compacted and for both dry and wet most compacted soils. The noise at the dry, loose sand soil stayed about the same, even with the blade electrodes. The overall noise level of the blades is more consistent than it was with the cylindrical electrode.

## 3.2. Considerations about the transversal effect

Even though the effect of vibration is expected to be the strongest in the direction of the vibration, we can expect VICOF in the transversal direction, too. One reason for that is that the laws of elastic deformations also have a transversal component. Another reason comes from the percolation effect of disordered systems. When the soil is loose, the current distribution follows a random percolation pattern with large spatial and directional fluctuations in the current density distribution. In this case, modulation of the microscopic resistance elements in the Y (longitudinal) direction can produce modulation of the macroscopic conductance measured in the X (transversal) direction. Figure 4 presents the simulation model and the ratio of $dR_{s,T}/R_s$ and $dR_{s,L}/R_s$ versus the soil bulk density.

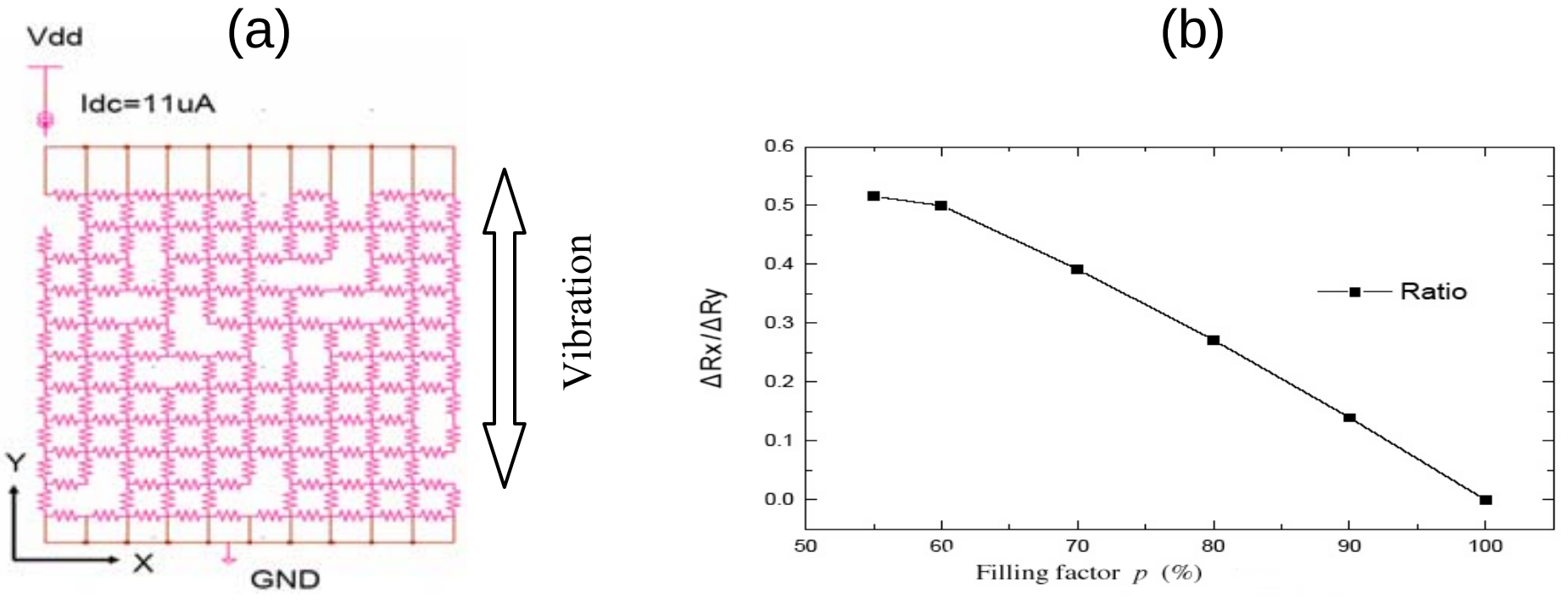

Fig.4. A simplified model study of the transversal versus the longitudinal VICOF effect. *(a)* The simulated two-dimensional square lattice of uniform resistors while measuring the longitudinal effect. *(b)* The ratio of the averaged transversal and longitudinal resistance changes when cutting out one resistor in the Y direction which is the direction of the vibration.

In this simple model, the transversal effect is zero when the Filling factor (*p*) is 100%, which is relevant for the most compacted soil. When the *p* is reduced toward the percolation limit (loose unconnected soil) the transversal effect is increasing and the transversal and longitudinal effects converge. Based on the computer simulation, the ratio of the two effects at various p values is shown in Figure 4 (b). When the *p* is decreasing from the limit of compacted soil toward the loose/unconnected soil limit, the transversal effect is increasing and the transversal and longitudinal effects converge.

The result shown in figure 4 (b) is independent of the value of the resistance of uniform resistors. Therefore, neither the vibration strength nor the conductivity of microscopic soil elements is expected to have an effect on data normalized in this way. Therefore we expect a reduced impact of salinity and water content when we use this normalization.

## 3.3. Normalized VICOF measurement and soil bulk density

Figure 5 presents the measurement results. As expected based on the theory, the normalized VICOF is inversely related to the bulk density for both longitudinal and transversal situation. This relationship is clearly seen and significant at $\alpha$=0.001 in case of dry natural sand with linear fit having $r^2$=0.7. On the other hand, in case of wet natural and salinized soils, the slope of correlation is also negative, but small for both transversal and longitudinal signals and response to bulk density (slope) is not significantly different from zero at $\alpha$=0.05.

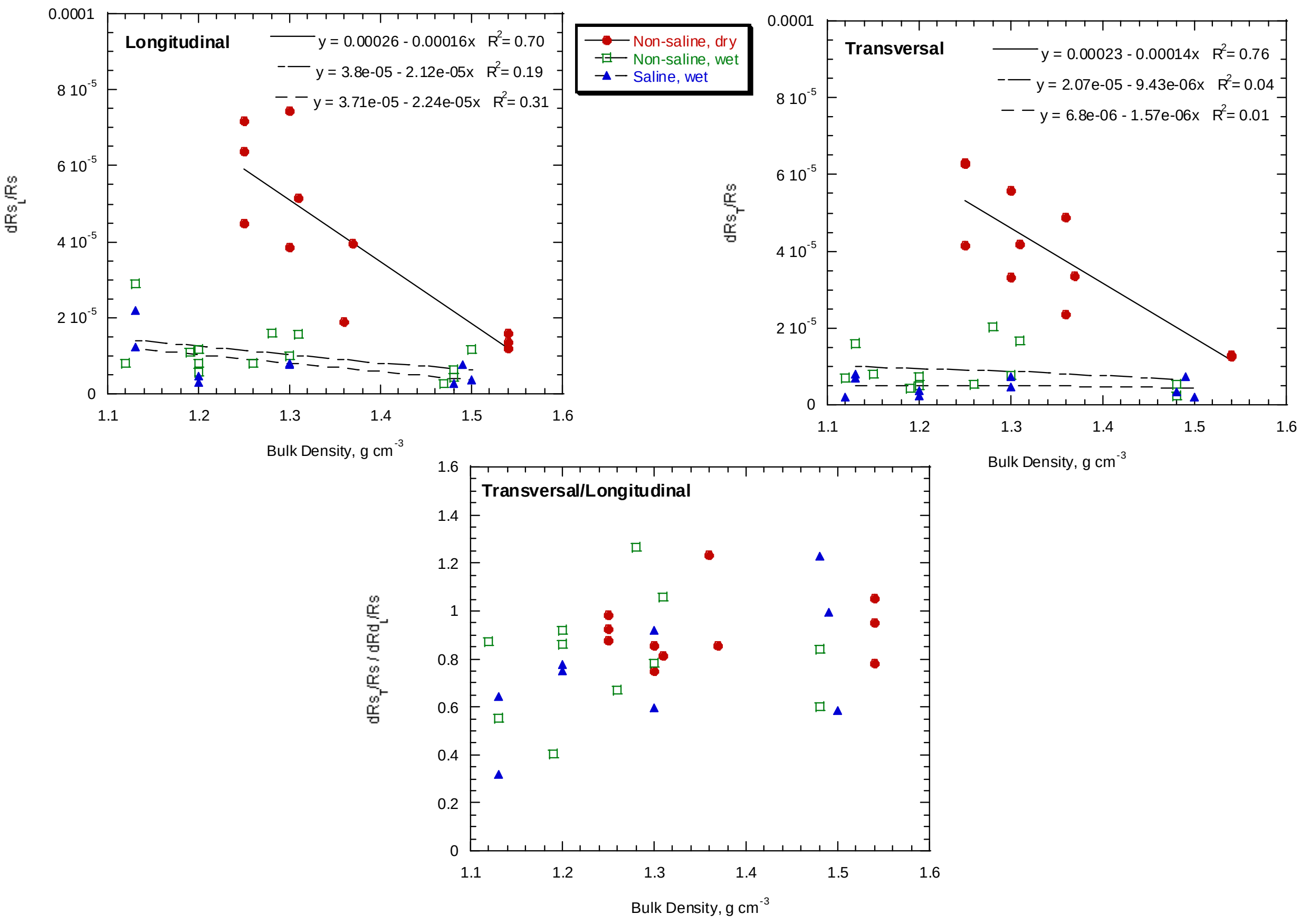

Fig. 5. Longitudinal, transversal normalized VICOF and the ratio of $dR_{s,T}/R_s$ and $dR_{s,L}/R_s$ versus soil bulk density.

Electrical conductivity measurements are very popular for measuring bulk properties of soils in soil science. However, salinity hinders many possible electrical conductivity applications. But normalized VICOF is not affected by strong salinity in wet soils. The tendency is similar in both longitudinal and transversal cases, although the measurement had weak response to bulk density. We speculate that the electrode material was not appropriate and too heavy to respond properly to vibration. The difference in material and dimensions in the cylindrical and blade electrodes may explain the smaller normalized VICOF response in smaller soil bulk densities than it was shown in the previous measurements [6]. Improved vibration method might also yield stronger signals.

The difference between the computer simulated and measured ratio originates from the difference in the transversal signals. The measured $dR_{s,T}/R_s$ showed larger fluctuation than expected even at larger bulk densities. This increase in fluctuation might be related to additional vibrations of the two electrodes in the transversal mode where the direction of the vibration is parallel to the contact plane thus the soil provides less support. Slight variations in the parallel positioning of blade electrodes done manually between alternating measurements may have introduced additional error. The effect of inhomogeneity within the soil sample needs also testing.

To improve our understanding of the underlying processes on how vibration affects the electrical conductivity, further studies are needed to relate the conductivity fluctuation modulated by vibration to the soil matrix, water, adsorbed cations and their exchange processes in the soil/liquid interface in microscopic scale.

## 4. CONCLUSIONS

1 Blade electrodes outperformed the cylindrical electrode based on the comparison of the set-up noise in the measurement of normalized vibration-induce conductivity (VICOF) in relation to bulk density and wetness of fine sand soils.

2 Theoretical considerations and a computer simulation were presented for explaining the inverse relationship of the longitudinal and transversal normalized VICOF to bulk density, as well as the ratio of transversal/longitudinal normalized VICOF. According to the theory, this ratio is independent of moisture and salinity affects.

3 On dry sand, the measurement results proved that there is an inverse correlation between the bulk density and normalized VICOF for both longitudinal and transversal cases. On wet non-saline and saline samples, the linear correlation coefficient was negative, but insignificantly different from zero ($\alpha$=0.05). Strong salinity did not affect the normalized VICOF signal in wet conditions.

4 The ratio of transversal and longitudinal normalized VICOF had approximately the same range for dry, wet and salinized soils. The main difference between the simulation model and measurement results was attributed to possible measurement circumstances, such as the electrode material, and the variation in the relative geometric arrangement of blades and of the direction of vibration. Improvements in these and in the vibration method are expected to improve the normalized VICOF signal, the ratio, and their response the soil bulk density. Fundamental studies of the vibration effect on the mechano-electrical soil responses can further the understanding and modeling of this process and its applications.

## ACKNOWLEDGEMENT

Our appreciation is expressed to Dr. Kevin McInnes for his suggestions in the soil compaction, and to Dr. Thomas Hallmark for guidelines in the salinization process. We thank Mr. Richard Epting for fabricating the blade electrodes and template used in the measurement. The research was supported by Texas Agricultural Experiment Station.

## REFERENCES

1. Grossman, R.B. and T.G. Reinch, "Bulk density and linear extensibility", Chapter 2.1. Methods of Soil Analysis. Part 4, Physical Methods, SSSA Book Series 5, 201-228 (2002).
2. Timm, L.C., L.F. Pires, K. Reichardt, R., R. Rovetti, J.C.M. Oliveira and O.O.S. Bacchi, "Soil bulk density evaluation by conventional and nuclear methods", *Australian J. of Soil Res.* 43, 97-103 (2005).
3. Yu, X. and V.P. Drnevich, "Soil water content and dry density by Time Domain Reflectometry", J. Geotech. And Geoenvir. Engrg.,130 (9) 922-934 (2004).
4. Kish, L.B., C.L.S. Morgan and A.Sz. Kishné, "Vibration-induced conductivity fluctuation (VICOF) testing of soils", *Fluctuation and Noise Letter*, 6(4), L359-L365 (2006).
5. Corwin, D.L. and S.M. Lesch, "Application of soil electrical conductivity to precision agriculture: Theory, principles, and guidelines", Agron. J. 95, 455-471 (2003).
6. Kishné, A.Sz., C.L.S. Morgan and L.B. Kish, "Measuring soil bulk density by using vibration-induced conductivity fluctuation", 18th World Congress of Soil Science, July 9-15, 2006, Philadelphia, PA, USA, Abstracts, p. 282.